\documentclass[twocolumn,apl,graphicx]{revtex4-1}
\usepackage{amssymb}
\usepackage{amsmath}
\usepackage{graphicx}
\usepackage{dcolumn}
\usepackage{bm}
\usepackage{color}

\input{tcilatex}

\begin{document}

\title{Magnetic moment manipulation by triplet Josephson current}
\author{N. G. Pugach}
\email{pugach@magn.ru}
\affiliation{Faculty of Physics, M.V. Lomonosov Moscow State University, GSP-2, Leninskie
Gory, 119991 Moscow, Russia}
\author{A. I. Buzdin}
\affiliation{Institut Universitaire de France and
University Bordeaux 1 LOMA, UMR 5798, F-33405 Talence, France}
\date{\today }

\begin{abstract}
The induced magnetic moment, provided by the bands electrons, is calculated
in a variety of Josephson junctions with multilayered ferromagnetic (F)
weak link. The noncollinear magnetization of the F layers provides the conditions necessary to generate triplet superconducting correlations. It leads
to the long-range induced magnetic moment, emerging in the
superconducting (S) layers. It is shown to be dependent on the Josephson
phase.
By tuning the Josephson current, one may control the long-range induced magnetic
moment. Alternatively, applying the voltage we can generate an oscillatory
magnetic moment. The detection of such a spin effect may serve as
independent evidence of the triplet superconductivity. The proposed
mechanism seems to be attractive for superconducting
spintronic devices with low dissipation.
\end{abstract}

\maketitle


The antagonistic nature of singlet superconductivity (S) and ferromagnetism (F)
makes their \ coexistence in bulk systems rather difficult \cite%
{FlBuzdin2002}. However a spatial separation of regions with the S and F
order avoids natural limitation and was used in many experimental
realizations of SF hybrid structures. The interplay of the superconducting
and ferromagnetic long-range orders in hybrid structures via the proximity
effect leads to a range of unusual physical phenomena \cite%
{BuzdinRev,VolkovRevModPhys,GKI}. For example, Josephson junction with a
ferromagnetic interlayer may have a spontaneous phase difference $\pi $ in
the ground state. Such $\pi $-junction have been successfully implemented to
strongly decrease the size of the single-flux-quantum circuits and reduce
noise in a superconducting qubit \cite{RyazanovNatPhys}.

Recently, a magnetic moment induced in the superconductor of a S/F bilayer
was reported \cite{Garifullin2009,Khaydukov2011} in accordance with the
previous predictions \cite%
{BVE:MagnMom:2004,KrivorucKoshina2002,Valls2004,Linder:MagnMom2009}
.
Another unusual effect highlighted in the SF heterostructures is a triplet
superconductivity. It was demonstrated theoretically \cite%
{BVE:PRB2001,VolkovRevModPhys,Shekhter}
that a non-collinear
magnetization in the SF heterostructures may lead to the creation of
spin-triplet superconducting correlations with a non-zero total spin
projection on the quantization axis. The exchange magnetic field does not
destroy them, thus leading to a long-range superconducting correlations
penetrating into the F region. The experimental evidence of such triplet
correlations (TC) was revealed by the recent observation of long-range
Josephson currents \cite{KeiserNature,Khaire,RobinsonScience}
that was
an important breakthrough in this domain. Interestingly, though at the
center of such Josephson junction the current is carried exclusively by the
triplet component, it does not produce magnetic moment. Nevertheless the
magnetic effect produced by the TC with the inherent nontrivial spin
structure of the Cooper pairs also seems to be very attractive for
applications. The challenging task is to control the magnetic moment of the
long-ranged TC to design new superconducting spintronics devices with
low dissipation.

Here we demonstrate under which conditions it may be possible to generate the
magnetic moment by the TC in the Josephson junction.
Such induced magnetization occurs at a relatively large distance and it is
sensitive to the superconducting phase difference. This opens
interesting perspectives to couple the Josephson effect with spintronics.

\begin{figure}[!tbh]
\begin{center}
\includegraphics{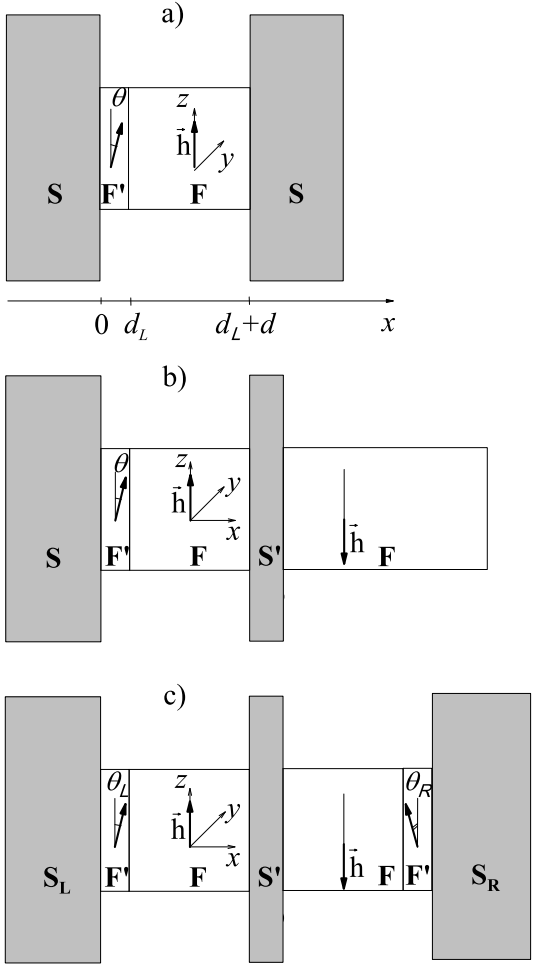}
\end{center}
\caption{The sketches of SF'FS Josephson junction (a), and more complicated
structures (b) SF'F$^{\uparrow }$S'F$^{\downarrow },$ and (c) SF'F$%
^{\uparrow}$S'F$^{\downarrow}$F'S.}
\label{Sketches}
\end{figure}

For simplicity, we consider below the diffusive limit and a temperature
close to the critical temperature $T_{c}$ which allows \ to describe the
underlying physics in the framework of the linearized Usadel equations. In
this limit the F and S coherence lengths are $\xi _{f}=\sqrt{D_{f}/h},$ $\xi
_{s}=\sqrt{D_{s}/2\pi T_{c}}$, where $D_{f,s}$ is the corresponding
diffusion coefficient, $h$ is the exchange field in the ferromagnet. 
To generate the
long-range TC it is necessary to have non-collinear magnetization or
spin-active interfaces. Here we concentrate on the case of a composite
non-collinear F layer. We start with a SF'FS structure (Fig.1(a)) with
semi-infinite S electrodes, and thicknesses $d_{L}$ and $d$ of the F' and F
layers respectively. The origin of the \textit{x}-axis, that is
perpendicular to the layer plane, locates at the left SF' interface. The
magnetization of the middle F layer is aligned along the \textit{z}-axis,
while the magnetization of the left F' layer is tilted by the angle $\theta $
from the \textit{z}-axis in the $yz$-plane (\ref{Sketches}). Naturally our results remain valid even if the magnetization of the F' layers are in the $%
xz$-plane. As it has been noted in \cite{HouzetBuzdin2007}, the
optimal condition for the TC generation corresponds to $d_{L}\lesssim \xi
_{f}\ll d$.

A SF heterostructure with a ferromagnet where an interface magnetization
rotates easier than the magnetization in the bulk may serve as a suitable
example. Note that it was found experimentally \cite%
{Clavero2009,Baczewski,Huttel} that a thin layer of vanadium in the vicinity
of an Fe interface shows different magnetization and coercitivity with
respect to that in the main F layer. A suitable structure for the
triplet weak link may be also a combination of 3-d ferromagnetic films of
different thicknesses or different compounds, having different anisotropy.
It was shown \cite{WeidesAnisotr} that the magnetic anisotropy of the Ni
layer changes from the perpendicular to the in-plane with an increase of the
layer thickness.
Another example could be the Ni-CuNi combination, where
the CuNi alloy exhibits a strong perpendicular anisotropy \cite{Vinnikov}.

Near $T_{c}$
the induced magnetic moment of the electrons (IMM) may be written as \cite%
{Eschrig:MagnMom2005}
\begin{equation}
\delta \mathbf{M}(x)=-4\mu _{B}N_{0}\pi T_{c}\sum\limits_{\omega >0}\mathrm{%
Im}f_{0}(x,\omega )\mathbf{f}^{\ast }(x,\omega ),  \label{dM}
\end{equation}%
where $N_{0}=N_{0f},N_{0s}$ is the density of states at the Fermi level in
the F or S metal. $f_{0}$ is a short-range singlet component, and $\mathbf{f}%
=(f_{x},f_{t},f_{z})$, where $f_{x}=0$, $f_{z}$ is short-range and $f_{t}$
is long-range triplet component (TC) of the anomalous Usadel function.
Consequently, only the $\delta M_{y}$ 
component of the induced magnetic moment may contain the long-range
correlations.

The characteristic value of the IMM in the superconductor can be easily
estimated from (\ref{dM}) to be of the order of $\mu _{B}N_{0}T_{c}$ \cite%
{Eschrig:MagnMom2005}. It is by a factor $T_{c}/h$ smaller than the
electrons polarization in the ferromagnet.

The anomalous functions obeys the linearized Usadel equations in the F and S
metals \cite{EschrigPRB2005}

\begin{eqnarray}
\left( -D_{f,s}\frac{\partial ^{2}}{\partial x^{2}}+2\omega \right)
f_{0}(x)+2i\mathbf{f}(x)\cdot \mathbf{h}(x) &=&2\Delta \exp \left( \pm i%
\frac{\varphi }{2}\right) ,  \notag \\
\left( -D_{f,s}\frac{\partial ^{2}}{\partial x^{2}}+2\omega \right) \mathbf{f%
}_{t}(x)+2if_{0}(x)\mathbf{h}(x) &=&0.  \label{UE}
\end{eqnarray}%
The exchange magnetic field $\mathbf{h}\ $\ is present only in the
ferromagnet, while the superconducting order parameter $\Delta \neq 0$ only
in the S layers, $\varphi $\ is the superconducting phase difference on the
junction, the sign "$+(-)$" is taken for the right (left) electrode. For $%
d_{L}\lesssim \xi _{f}$, we may use the Taylor expansion for the Usadel
function $f(x)$ in the thin left F' layer.
In the F layer the function may be written in terms of its values at the
interfaces \cite{HouzetBuzdin2007} $f_{i}(x)=f_{i}(d_{L})\sinh
q_{i}(d_{L}+d-x)/\sinh q_{i}(d)+f_{i}(d_{L}+d)\sinh q_{i}(x-d_{L})/\sinh
q_{i}(d)$,
where $i=+,-,t$, and $f_{\pm }(x)=f_{0}(x)\pm f_{z}(x),$ $q_{t}\equiv q_{0}=%
\sqrt{2\omega /D_{f}},$ $q_{\pm }=\sqrt{2(\omega \pm ih)/D_{f}}\approx (1\pm
i)/\xi _{f}$ at $h\gg T_{c}$.
For simplicity, we assume the boundary resistance at the F'F interface is negligible and so we treat it as being zero, a thin normal metal interlayer 
influences insignificantly, the boundary resistance
of SF interfaces is also zero, and so the anomalous Usadel function satisfy the
boundary conditions \cite{KL}:

\begin{equation}
\left. f(x)\right\vert _{s}=\left. f(x)\right\vert _{f},\left. \frac{%
\partial f}{\partial y}\right\vert _{s}=\gamma \left. \frac{\partial f}{%
\partial y}\right\vert _{f},\gamma =\frac{\sigma _{f}}{\sigma _{s}},
\label{RBC}
\end{equation}%
where $\sigma _{s},\sigma _{f}$ \ are the normal state conductivities of the S and
F layers. The inequality $\gamma \ll \frac{\xi _{f}}{\xi _{s}}$ provides
so-called rigid boundary conditions, which allows to neglect the
suppression of superconductivity in the S layer due to the proximity effect%
, thus $\Delta =const$. We may also use $\Delta =const$
when the S electrodes are much wider in the $y$ or $z$ direction than the F
link (Fig.1), if the F link is  narrower than $\xi _{s}$.

The rigid boundary conditions yield:
\begin{eqnarray}
f_{0}(d_{L}+d) &=& \frac{\Delta }{\omega }\exp (i\varphi /2),  \notag \\
f_{t}(d_{L}+d) &=& \gamma \frac{\xi _{s}}{\xi _{0}} f_{t}(d_{L})/ \sinh
q_{0}d, \\
f_{t}(d_{L}) = &-& i\frac{d_{L}^{2}}{\xi _{f}^{2}}\sin (\theta) \frac{\Delta
}{\omega }\exp (-i\frac{\varphi}{2}) .  \notag
\end{eqnarray}
This leads to suppression of the
TC near the FS boundary due to the small parameter $\gamma\frac{ \xi _{s}}{\xi _{0}}=%
\frac{N_{0f}}{N_{0s}}\sqrt{\frac{D_{f}}{D_{s}}}\ll 1$.
Here $\xi _{0}=\sqrt{D_{f}/2\pi T_{c}}$ is the normal coherence length.

\begin{figure}[tbh!]
\begin{center}
\includegraphics{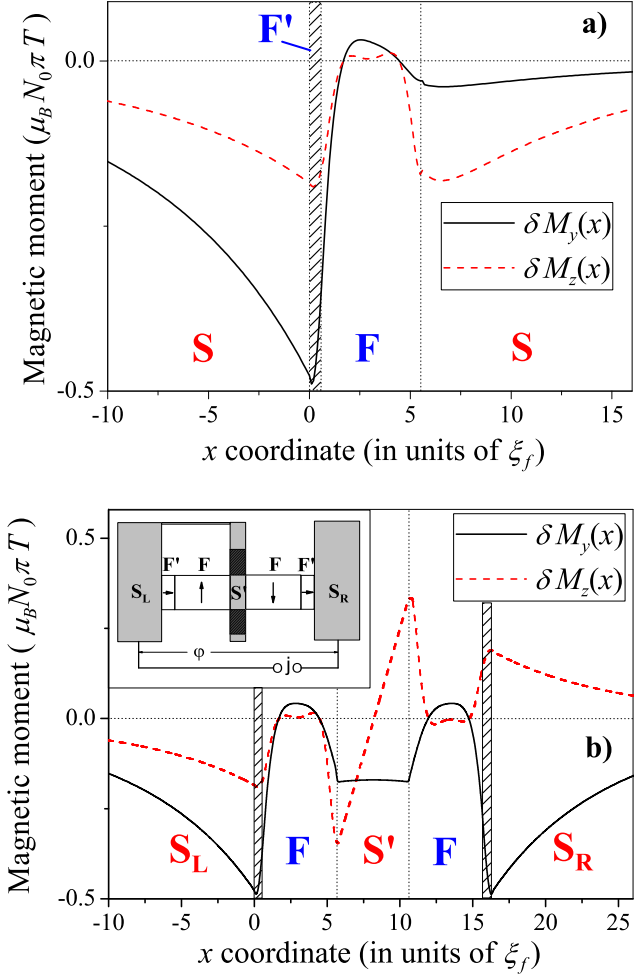}
\end{center}
\caption{(Color online) The induced magnetic moment distribution along (a) the SF'FS
junction (fig.1(a)), and (b) the structure in Fig.1(c). The inset shows the possible
incorporation of the structure (Fig.1(c)) into a superconducting circuit.
$\protect\xi _{0}=5\protect\xi _{f},\protect\xi _{s}=10\protect%
\xi _{f},$ the thicknesses of layers $d_{L}=d_R=0.7\protect\xi _{f},d=5\protect%
\xi _{f},$ Josephson current is absent $\varphi =\varphi_L=\varphi_R=0$ %
and $\mathrm{Im}f_{0}=0$. $\theta =\theta_L=\theta_R=\pi /2$, $T\approx 0.8T_{c}$. The FS
boundary parameter $\protect\gamma =0.05$.}
\label{SFFS}
\end{figure}

Thus, we obtain the expression for the $y$ component of the IMM in the right
S layer

\begin{eqnarray}
&&\delta M_{y}(x)=-4\mu _{B}N_{0s}\pi T\frac{d_{L}^{2}\sin \theta }{\xi
_{f}^{2}}\gamma \frac{\xi _{s}}{\xi _{0}}\cos \varphi \times   \notag
\label{dMy} \\
&&\times \sum\limits_{\omega >0}\frac{\Delta ^{2}}{\omega ^{2}}\frac{\exp
[-q_{s}(x-d_{L}-d)]}{\sinh q_{0}d},
\end{eqnarray}%
where $q_{s}=\sqrt{2\omega /D_{s}}$. If $d\lesssim \xi _{0},d_{L}\sim \xi
_{f}$, $\gamma \xi _{s}/\xi _{f}\lesssim 1$, the IMM (\ref{dMy}) may be
estimated at the FS boundary (at $T$ $\lesssim 0.5 T_{c}$) as $\delta M_{y}\sim
-\mu _{B}N_{0s}T_{c}\frac{\xi _{f}}{d}\sin \theta \cos \varphi $.

At the center of the SF'FF"S structure, considered in the work \cite%
{HouzetBuzdin2007}, we have the triplet supercurrent but the magnetization
is zero because the singlet component vanishes. This component is necessary
to generate the induced electron's magnetization (\ref{dM}).

The IMM has a maximum value near the FS interfaces. It decreases and
oscillates at the distance $\xi _{f}$ in the ferromagnet and falls down at
the distance  $\xi _{s}$ in the superconductor in the considered SF'FS
structure (Fig.2(a)).
$\xi _{s}$ is a characteristic scale of change of the anomalous Usadel
function in a superconductor, therefore, the induced magnetic moment may be
detected at a distance $\lesssim \xi _{s}$ from the junction in the $yz$
plane as well.
The IMM (\ref{dMy}) depends on the Josephson phase difference,
because the triplet and the singlet components are generated by
different S electrodes.
The $\delta M_{y}$ at $\varphi =0$ near the FS interface has opposite
direction to $M_{y}$ in the F' layer. It reminds the situation with the $%
\delta M_{z}$, that appears as a result of the short-range proximity effect
near the FS interface and has a direction opposite to the F layer
magnetization \cite{BVE:MagnMom:2004}.
At the FS
interface this component is phase independent and may be estimated as $%
\delta M_{z}\sim -\mu _{B}N_{0s}\pi \frac{\Delta ^{2}}{2T}\gamma \frac{\xi
_{s}}{\xi _{f}}$, that corresponds to the short-range IMM
calculated in the FSF sandwich \cite{Eschrig:MagnMom2005}.


The best way to isolate the phase sensitive $\delta M_{y}$ component seems
to be to consider the structure Fig.1(b) with the thin S' layer, sandwiched
between two antiparallel ferromagnets. Due to the small thickness of the S'
layer $d_{s}\ll \xi _{s}$ the averaged induced $\delta M_{z}$ component will
be zero (Fig.2(b)). Calculating the triplet Usadel function at the S' layer we  find
\begin{eqnarray}
f_{t}(d_{L}+d) &=&-i\frac{d_{L}^{2}}{\xi _{f}^{2}}\sin (\theta )\frac{\Delta
}{\omega }\exp (-i\varphi /2)\times  \\
&&\times \frac{\gamma \xi _{s}/\xi _{0}}{q_{s}d_{s}\sinh q_{0}d+(\gamma \xi
_{s}/\xi _{0})\exp q_{0}d}.  \notag
\end{eqnarray}

The amplitude $f_{t}$ is maximal for $d_{s}\rightarrow 0$. However, the
condition of the superconductivity existence in the S' layer imply $%
d_{s}>\gamma \xi _{s}^{2}/\xi _{f}$ \cite{BuzdinRev}. Nevertheless, this
restriction may be easily overcome if the S' electrode exceeds the lateral dimension
 of F layers in the $yz$ plane. In this situation the magnetic moment
will be induced at the periphery of the S' layer where its inherent singlet
superconductivity will coexist with the induced TC. The IMM in the S' layer
(if $d_{s}>\gamma \xi _{s}^{2}/\xi _{f}$) writes

\begin{equation}
\delta M_{y}(x)\sim -4\mu _{B}N_{0}\pi T\cos \varphi \frac{d_{L}^{2}\sin
\theta }{\xi _{f}^{2}}\sum\limits_{\omega >0}\frac{\Delta ^{2}}{\omega ^{2}}%
\frac{\gamma \xi _{s}/\xi _{0}}{q_{s}d_{s}\sinh q_{0}d}.
\end{equation}%

The Josephson junction corresponding to the setup (b) in Fig.1 should have
extremely low critical current which complicates the in-situ phase control.
The triplet Josephson junction presented in Fig.1(c) would have much larger
critical current due to the interference between two sources of TC \cite%
{HouzetBuzdin2007}. Let's consider the
structure (Fig.1(c)) with the left and the right F' layers with
magnetization aligned at the angles $\theta _{L}$ and $\theta _{R}$ to $OZ$
axes, and thicknesses $d_{L}$ and $d_{R}$ respectively, the F
layers have the same thickness $d$.
If this structure is a part of the multiterminal device it is possible to
have the different phases $-\varphi _{L}$, $0$, and $+\varphi _{R}$
respectively at the left, middle, and the right S electrodes.
If the F layers have opposite magnetization and
the parameters the F' layers are equal
$d_L=d_R, \theta_L=\theta_R, \varphi_L=\varphi_R$,
the $\delta M_{z}$ component is antisymmetric relative to the center
 of the S' layer providing its zero average value,
 while the $\delta M_{y}$ component is symmetric (Fig.2(b)), and may be written
 in the S' layer as
\begin{eqnarray}
&&\delta M_{y}(x)=-4\mu _{B}N_{0}\pi T\times  \\
&&\times \sum\limits_{\omega >0}\frac{\Delta ^{2}}{\omega ^{2}}\frac{\gamma
\xi _{s}/\xi _{0}\left[ d_{R}^{2}\sin \theta _{R}\cos \varphi
_{R}+d_{L}^{2}\sin \theta _{L}\cos \varphi _{L}\right] }{\xi _{f}^{2}\sinh
q_{s}d_{s}\sinh q_{0}d}.  \notag
\end{eqnarray}%
It contains the same attenuation parameter $\gamma \xi _{s}/\xi _{0}$. Note
that the most suitable for the triplet magnetic moment generation should be
the system with a very thin middle S' layer $d_{s}\ll \gamma \xi
_{s}^{2}/\xi _{0}$. In this case the attenuation of the TC vanishes but the
superconductivity will be destroyed in the region between the F layers.
However, if the lateral size of the S' layer exceeds that of the F layers,
the singlet superconductivity will interfere with TC at the lateral distance
$\xi _{s}$ from the boundary of the F layer (shaded regions in Fig.2(inset)). In this region we have
the optimal conditions for the IMM observation, for the structure
(Fig.1(c)) at $d_L=d_R\sim \xi_f$ and $d\sim \xi_0$
the TC magnetization
$\delta M_{y}\sim -\mu _{B}N_{0} T_c\left[ \sin \theta _{R}\cos \varphi_{R}+\sin \theta _{L}\cos \varphi _{L}\right] $.
The exterior S electrodes may be considered as a source of the
superconducting correlation, which forms the TC
at the F'F boundaries, and penetrates at a long distance through the
ferromagnet. Then the middle S' layer  serves as a detector of the
long-range induced magnetic moment (IMM).
Let us consider for illustration the setup presented in the inset of Fig.2.
 Here the superconducting electrodes $\mathrm{S_{L}}$ and S'
 have the same superconducting phase but the
superconducting current is flowing through the triplet Josephson junction
$\mathrm{S_{L}}$ - $\mathrm{S_{R}}$, the critical current of the S' -
$\mathrm{S_{R}}$ junction being vanishingly small (because in this
junction only a short ranged proximity effect is possible). In such a case
at the optimum conditions the IMM at the  S' electrode
should be $\delta M_{y}\sim \mu _{B}N_{0}T_{c}\cos \varphi $, where
$\varphi $ is a phase difference on the junction.

Changing the applied Josephson phase or the Josephson current
through the junction one may vary $\delta M_{y}(\varphi )$ at fixed
magnetization of the layers. In particular, applying the voltage to the
Josephson junction, we create a situation, when the phase oscillates in time
that results in the oscillations of the electron's magnetization, coupling
the magnetic dynamics with the superconducting one. If some part of the S'
electrode contains the magnetic atoms, the TC magnetization should polarize
them. Assuming the typical value of the exchange interaction between electron's
 spin and the localized moment $I\sim 10^{3}$ K and $T_{c}\sim 10$%
K, we may estimate that the TC polarization should be equivalent to
the magnetic field $\sim 1-10$ kOe.

Penetrating in other ferromagnet being in a contact with the
S' layer, the IMM may operate its magnetization.
The IMM tuned by the Josephson current may be used in spintronic devices
instead of the spin-torque effect, which needs a significant dissipative
current. Moreover, this effect may be used for operating the new
types of the magnetic Josephson valve and memory \cite{RyazanovMagnMem},
where the switch of the magnetization of a soft magnetic weak link
changes the critical current of the readout Josephson junction.

In conclusion we have demonstrated that the superconducting
triplet correlations can generate a magnetic moment,
sensitive to the superconducting phase. In the Josephson junction
with the composite non-collinear ferromagnetic interlayer
this mechanism provides a direct coupling
between the superconducting current and magnetization.

\begin{acknowledgments}
We are grateful to J. Robinson for valuable discussions and remarks.
The work was supported by
the Russian Foundation for Basic Research,
the program of LEA Physique Theorique et Matiere Condensee,
and the European IRSES program SIMTECH.
\end{acknowledgments}


\begin{thebibliography}{99}
\bibitem{FlBuzdin2002} J. Flouquet and A. Buzdin, Phys. World \textbf{15},
41 (2002).

\bibitem{BuzdinRev} A. I. Buzdin, Rev. Mod. Phys. \textbf{77}, 935 (2005).

\bibitem{VolkovRevModPhys} F.S. Bergeret, A.F. Volkov, and K.B. Efetov, Rev.
Mod. Phys. \textbf{77}, 1321 (2005).

\bibitem{GKI} A. A. Golubov, M. Yu. Kupriyanov, E. I. Il'ichev Rev. Mod.
Phys. \textbf{76}, 411 (2004).

\bibitem{RyazanovNatPhys} A.K. Feofanov, Nature Phys. \textbf{6}, 593 (2010).

\bibitem{Garifullin2009} L.R. Salikhov, I.A. Garifullin, N.N. Garifyanov, L.
R. Tagirov, K. Theis-Brohl, K. Westerholt, and H. Zabel, Phys. Rev. Lett.
102, 087003 (2009), Phys. Rev. B 80, 214523 (2009).

\bibitem{Khaydukov2011} Yu.N. Khaydukov, V.L. Aksenov, Yu.V. Nikitenko, K.N.
Zhernenkov, B. Nagy, A. Teichert, R. Steitz, A. Ruhm, L. Bottyan, J.
Supercond. Nov. Magn. 24, 961 (2011).

\bibitem{BVE:MagnMom:2004} F.S. Bergeret, A.F. Volkov, K.B. Efetov, Phys.
Rev. B \textbf{69}, 174504 (2004).

\bibitem{KrivorucKoshina2002} V.N. Krivoruchko and E.A. Koshina Phys. Rev. B
\textbf{66}, 014521 (2002).

\bibitem{Valls2004} K. Halterman, O. T. Valls, Phys. Rev. B \textbf{69},
014517 (2004).

\bibitem{Linder:MagnMom2009} J. Linder, T. Yokoyama and A. Sudbo, Phys. Rev.
B \textbf{79}, 054523 (2009).

\bibitem{BVE:PRB2001} F.S. Bergeret, A.F. Volkov, and K.B. Efetov, Phys.
Rev. B \textbf{64}, 134506 (2001).

\bibitem{Shekhter} A. Kadigrobov, R.\ I. Shekhter, and M. Jonson, Europhys.
Lett. 54, 394 (2001).

\bibitem{KeiserNature} R.S. Keiser, S. T. B. Goennenwein, T. M. Klapwijk, G.
Miao, G. Xiao, and A. Gupta , Nature \textbf{439}, 825 (2006).

\bibitem{Khaire} T.S. Khaire W. P. Pratt, Jr., and N. O. Birge, Phys. Rev.
Lett. \textbf{104}, 137002 (2010).

\bibitem{RobinsonScience} J.W.A. Robinson, J.D.S. Witt, M.G. Blamire,
Science \textbf{329}, 59 (2010).

\bibitem{HouzetBuzdin2007} M. Houzet, A.I. Buzdin, Phys. Rev. B \textbf{76},
060504 (2007).

\bibitem{Baczewski} L. T. Baczewski, P. Pankowski, A. Wawro, K. Mergia, and
S. Messoloras, Phys. Rev. B \textbf{74}, 075417 (2006).

\bibitem{Huttel} Y. Huttel, G. van der Laan, T. K. Johal, N. D. Telling, and
P. Bencok, Phys. Rev. B \textbf{68}, 174405 (2003).

\bibitem{Clavero2009} C. Clavero, J. R. Skuza, Y. Choi, Phys. Rev. B \textbf{%
80}, 024418 (2009).

\bibitem{WeidesAnisotr} M. Weides, Appl. Phys. Lett. \textbf{93}, 052502
(2008).

\bibitem{Vinnikov} I. S. Veschunov, V. A. Oboznov, A. N. Rossolenko, A. S.
Prokofiev, L. Ya. Vinnikov, A. Yu. Rusanov, D. V. Matveev, Pis'ma v ZhETF
\textbf{88}, 873 (2008).

\bibitem{Eschrig:MagnMom2005} T. Lofwander, T. Champel, J. Durst, and M.
Eschrig, Phys. Rev. Lett. \textbf{95}, 187003 (2005).

\bibitem{EschrigPRB2005} T. Champel, M. Eschrig Phys. Rev. B \textbf{72},
054523(2005).

\bibitem{KL} M. Yu. Kuprianov and V. F. Lukichev, Zh. Eksp. Teor. Fiz.
\textbf{94}, 139 (1988) [Sov. Phys. J.~Exp.~Theor.~Phys. \textbf{67}, 1163
(1988)].

\bibitem{RyazanovMagnMem} V. V. Bolginov, V. S. Stolyarov, D. S. Sobanin, A.
L. Karpovich, V. V. Ryazanov, JETP Lett. \textbf{95}, 366 (2012)  [Pis'ma v
ZhETF \textbf{95}, 408-413 (2012)].

\end{thebibliography}
\end{document}